\documentclass[aps,pre,twocolumn,showpacs]{revtex4}

\usepackage{graphicx}

\begin{document}

\title{Compression algorithm for multi-determinant wave functions}

\author{Gihan L.\ Weerasinghe}
\author{Pablo L\'opez R\'{\i}os}
\author{Richard J.\ Needs}
\affiliation{Theory of Condensed Matter Group,
             Cavendish Laboratory,
             J J Thomson Avenue,
             Cambridge CB3 0HE,
             United Kingdom}

\date{\today}

\begin{abstract}
A compression algorithm is introduced for multi-determinant wave
functions which can greatly reduce the number of determinants that
need to be evaluated in quantum Monte Carlo calculations.
We have devised an algorithm with three levels of compression, the
least costly of which yields excellent results in polynomial time.
We demonstrate the usefulness of the compression algorithm for evaluating
multi-determinant wave functions in quantum Monte Carlo
calculations, whose computational cost is reduced by factors of
between 1.885(3) and 25.23(4) for the examples studied.
We have found evidence of sub-linear scaling of quantum Monte Carlo
calculations with the number of determinants when the compression
algorithm is used.
\end{abstract}


\maketitle

\section{Introduction}
\label{sec:intro}

The variational and diffusion \cite{ceperley-1980,qmcrmp} quantum
Monte Carlo (VMC and DMC) methods are the most accurate known for
computing the energies of large numbers of interacting quantum
particles.
The crucial ingredient is an approximate trial wave function which
should be easy to evaluate, while giving a good approximation to the
true many-body wave function.
A standard approach is to use a Slater-Jastrow trial wave function
which consists of the product of determinants for the up and down-spin
electrons, multiplied by a Jastrow factor that describes dynamical
correlation \cite{DTN_jastrow,jastrow}.
We omit the Jastrow factor in the rest of this paper for conciseness.
Static correlation can be included by replacing the determinant
product by a multi-determinant expansion,
\begin{equation}
\label{eq:sj_mdet}
\Psi_{\rm MD}({\bf R}) =
  \sum_{k=1}^{N_s} c_k
    \Phi^\uparrow_k({\bf R}_{\uparrow})
    \Phi^\downarrow_k({\bf R}_{\downarrow}) \;,
\end{equation}
where $\Phi^\uparrow_k({\bf R}_{\uparrow}) =
\det\left[\phi_{a_{i,k}^\uparrow}({\bf r}_j^\uparrow)\right]$ and
$\Phi^\downarrow_k({\bf R}_{\downarrow}) =
\det\left[\phi_{a_{i,k}^\downarrow} ({\bf r}_j^\downarrow)\right]$
are determinants of up- and down-spin single-particle orbitals,
$a_{i,k}^\sigma$ is an index that selects the orbital which occurs in
the $i$th row of the $\sigma$-spin determinant in the $k$th term of
the expansion, ${\bf R}_\sigma$ denotes the set of $\sigma$-spin
electron coordinates, ${\bf R}=\{{\bf R}_{\uparrow},
{\bf R}_{\downarrow}\}$, and $c_k$ is the coefficient of the $k$th
term in the expansion.
The accuracy of $\Psi$ can be further improved by, for example,
increasing the number of terms in the expansion, $N_s$, or by
introducing a backflow transformation \cite{bf-ne,backflow}.

VMC and DMC calculations are normally performed by displacing
electrons one at a time, because this has been shown to be the most
efficient way to decorrelate consecutive electronic configurations
\cite{backflow,strategies}.
The displacement of a single electron requires the calculation of the
ratio of the wave functions at the new and old coordinates,
$\Psi({\bf R}^{\prime})/\Psi({\bf R})$.
In a standard single-determinant calculation this requires the
replacement of a single row of the Slater matrix by the vector of the
orbitals at the new position ${\bf R}^{\prime}$, and the required
calculation is performed using the Sherman-Morrison formula
\cite{qmcrmp}.
In a multi-determinant calculation $N_s$ such calculations must be
performed.
In a backflow calculation each electronic coordinate in the Slater
part of the wave function is replaced by a ``quasiparticle
coordinate'', which depends on all of the electronic coordinates, so
that each entry within each Slater matrix must be recalculated when
an electron is displaced, and its determinant must be reevaluated,
which is achieved using standard LU decomposition.
In a multi-determinant backflow calculation, $N_s$ Slater determinants
for each spin must be constructed and evaluated.

The repeated evaluation of the trial wave function and its first two
derivatives for different electronic coordinates ${\bf R}$ is
the main contribution to the computational cost of a QMC calculation,
which is, as discussed above, approximately proportional to $N_s$.
Methods to reduce the cost of evaluating multi-determinant wave
functions during QMC calculations have been developed in previous
studies \cite{clark_2011,nukala_2009}.
In this paper we introduce a determinant compression algorithm which
can significantly reduce the cost of evaluating a multi-determinant
trial wave function by reducing the number of determinants in the
expansion, by as much as a factor of 26.57 in the examples presented
here.
We have used the \textsc{casino} code \cite{casino} for the QMC
calculations reported here.

\section{Methodology}
\label{sec:method}

Quantum chemistry methods are often used to provide appropriate
multi-determinant wave functions for electronic systems.
In practice, the determinants contain $M_s$ distinct orbitals, and
different determinants often differ by only a single orbital.
Moreover, quantum chemistry methods group determinants into
configuration state functions (CSFs), and different CSFs may contain
the same determinant product $\Phi^\uparrow_k \Phi^\downarrow_k$.
The compression method we present here exploits these two facts.

\subsection{Basic determinant operations}
\label{sec:combine}

\subsubsection{Identical determinants}
\label{sec:dedup}

To achieve greater efficiency we combine repeated determinant products
so that each of them need only be evaluated once at each ${\bf R}$,
which is equivalent to simply adding together the terms from
identical determinant products, \textit{i.e.},
\begin{equation}
c_1 \Phi^\uparrow \Phi^\downarrow +
c_2 \Phi^\uparrow \Phi^\downarrow =
(c_1+c_2) \Phi^\uparrow \Phi^\downarrow =
c_1^\prime \Phi^\uparrow \Phi^\downarrow \;,
\end{equation}
where $c_1^\prime=c_1+c_2$ is the coefficient of the term arising from
the combination of two repeated determinant products.

We refer to this procedure as ``de-duplication'', which is the first
operation in our compression algorithm, and its computational cost
scales as $\mathcal{O}({N_s}^2)$, since all pairs of determinants
need to be compared to determine if they are equal.
The number of determinants in the resulting expansion is $N_d\leq
N_s$, while the number of distinct orbitals $M_d$ is equal to $M_s$.
Later stages of the compression algorithm can be simplified based on
the assumption that the expansion has been de-duplicated.

\subsubsection{Determinants differing by a single orbital}
\label{sec:compress_core}

It is convenient to express each Slater determinant in
Eq.\ (\ref{eq:sj_mdet}) using a compact vector notation consisting
of the list of orbitals that the determinant contains,
\begin{equation}
\label{eq:detmap}
\left[ \phi_{a_1}, \phi_{a_2}, \ldots, \phi_{a_n} \right] \equiv
\left|
\begin{array}{cccc}
  \phi_{a_1}({\bf r}_1) & \phi_{a_2}({\bf r}_1) & \cdots &
                          \phi_{a_n}({\bf r}_1) \\
  \phi_{a_1}({\bf r}_2) & \phi_{a_2}({\bf r}_2) & \cdots &
                          \phi_{a_n}({\bf r}_2) \\
  \vdots                & \vdots                & \ddots &
                          \vdots                \\
  \phi_{a_1}({\bf r}_n) & \phi_{a_2}({\bf r}_n) & \cdots &
                          \phi_{a_n}({\bf r}_n)
\end{array}
\right| \;.
\end{equation}
Central to the algorithm is an elementary identity from linear algebra
which allows two determinants to be combined if they differ by a
single row or column.
This is applicable to a multi-determinant expansion for terms where
the determinants of one spin type are equal and can be factored
out, and the determinants of the other spin type differ by a single
column, \textit{e.g.},
\begin{eqnarray}
\label{eq:example_detcombine1}
  c^\prime_1 [\phi_{a_1}, \phi_{a_3}, \ldots, \phi_{a_4}] \Phi^\downarrow
+ c^\prime_2 [\phi_{a_2}, \phi_{a_3}, \ldots, \phi_{a_4}] \Phi^\downarrow
  \nonumber \\
= [c^\prime_1 \phi_{a_1} + c^\prime_2 \phi_{a_2}, \phi_{a_3}, \ldots,
  \phi_{a_4}] \Phi^\downarrow
  \nonumber \\
= {\tilde c}_1 [{\tilde \phi}_{{\tilde a}_1}, \phi_{a_3}, \ldots, \phi_{a_4}]
  \Phi^\downarrow
 \;,
\end{eqnarray}
where ${\tilde \phi}_{{\tilde a}_1}$ is a new orbital resulting from
the linear combination of two of the original orbitals and ${\tilde
c}_1$ is a new coefficient, satisfying
\begin{equation}
 \label{eq:example_detcombine2}
 {\tilde c}_1 {\tilde \phi}_{{\tilde a}_1} =
 c^\prime_1 \phi_{a_1} + c^\prime_2 \phi_{a_2} \;.
\end{equation}
This operation can be applied to sets of more than two determinant
products, provided they all differ in the same column of the same-spin
determinant.

The compression algorithm we have developed applies this basic
operation repeatedly to all possible sets of determinants.
However, there may be multiple mutually-exclusive ways of combining
the determinants, and the size of the resulting expansion depends on
the choice of operations.
We discuss this in Sec.\ \ref{sec:group_terms}.

\subsection{Representation of compressed expansions}
\label{sec:represent}

A compressed multi-determinant expansion is of the form
\begin{equation}
\label{eq:compressed_exp}
\Psi_{\rm MD}({\bf R}) =
\sum_{k=1}^{N_c} {\tilde c}_k
 \det\left[ {\tilde \phi}_{{\tilde a}^\uparrow_{ik}}
            ({\bf r}^\uparrow_j) \right]
 \det\left[ {\tilde \phi}_{{\tilde a}^\downarrow_{ik}}
            ({\bf r}^\downarrow_j) \right] \;,
\end{equation}
where $N_c$ is the number of terms in the compressed expansion,
${\tilde c}$ are the compressed expansion coefficients,
${\tilde \phi}$ are the compressed orbitals, and
${\tilde a}^\sigma_{ik}$ is an index that selects which
compressed orbital occurs in the $i$th row of the $\sigma$-spin
determinant in the $k$th term of the compressed expansion.
The expansion coefficients are
\begin{equation}
\label{eq:compressed_coeff}
{\tilde c}_k =
  \pm c^\prime_{\nu_{k1}}
  \prod_{p=2}^{P_k}
  \frac{c^\prime_{\nu_{kp}}}
       {c^\prime_{\delta_{kp}}} \;,
\end{equation}
and the compressed orbitals are
\begin{equation}
\label{eq:compressed_orb}
{\tilde \phi}_a ({\bf r}) =
  \sum_{x=1}^{X_a}
  \pm
  \prod_{q=1}^{Q_{ax}}
  \frac{c^\prime_{n_{axq}}}
       {c^\prime_{d_{axq}}}
  \phi_{\mu_{ax}} ({\bf r}) \;,
\end{equation}
where $P$, $X$, and $Q$ are sum and product lengths, and $\nu$,
$\delta$, $n$, $d$, and $\mu$ are indices, all of which arise from the
application of the compression operations described above.
The $\pm$ signs account for any required row exchanges in the
determinants.
In this notation, the compression operation exemplified in
Eq.\ (\ref{eq:example_detcombine1}) is such that
\begin{equation}
{\tilde c}_1 = + c^\prime_1 \;,
\end{equation}
and
\begin{equation}
{\tilde \phi}_{{\tilde a}_1} ({\bf r}) =
 + \phi_{a_1}({\bf r})
 + \frac{c^\prime_2}{c^\prime_1} \phi_{a_2}({\bf r}) \;,
\end{equation}
which satisfies Eq.\ (\ref{eq:example_detcombine2}), as required.
Therefore in this case $P_1=1$, $\nu_{1,1}=1$, $X_{{\tilde a}_1}=2$,
$Q_{{\tilde a}_1,1}=0$, $\mu_{{\tilde a}_1,1}=a_1$,
$Q_{{\tilde a}_1,2}=1$, $n_{{\tilde a}_1,2,1}=2$,
$d_{{\tilde a}_1,2,1}=1$, and $\mu_{{\tilde a}_1,2}=a_2$.

A compressed expansion is fully determined by specifying
${\tilde a}$, $P$, $X$, $Q$, $\nu$, $\delta$, $n$, $d$, $\mu$, and
the signs in Eqs.\ (\ref{eq:compressed_coeff}) and
(\ref{eq:compressed_orb}).
Expressing the compressed expansion in this manner is useful because
the orbitals and coefficients can be quickly reconstructed using
Eqs.\ (\ref{eq:compressed_coeff}) and (\ref{eq:compressed_orb}) when
the original expansion coefficients, the $\{c_k\}$ of
Eq.\ (\ref{eq:sj_mdet}), change, as is the case during wave
function optimization within QMC.

\subsection{Choosing the optimal set of operations}
\label{sec:group_terms}

It is convenient to express the principles of the compression
algorithm using set theory notation.
Let $P$ be a set whose elements are the terms in the
de-duplicated multi-determinant expansion, $P = \{p_k \equiv
c^\prime_k \Phi_k^\uparrow \Phi_k^\downarrow\}$, of size $|P|=N_d$.
Let $u_i$ be a subset of $P$ such that its elements can be combined
via the compression operation of Sec.\ \ref{sec:compress_core}, and
$U$ the set of all possible such sets, $U=\{u_i\}$.
Note that $u_i$ is allowed to contain only one term, and that any
two elements $u_i$ and $u_j$ may contain the same term $p_k$.

A valid compressed expansion can be obtained by finding a subset
$V=\{v_i\}$ of $U$ that satisfies the conditions that (a) $V$ contains
all terms in $P$, $\bigcup_i v_i = P$, and (b) each term in $P$ is
contained in only one element of $V$, $v_i \cap v_j = \emptyset
~~ \forall~ i \neq j$.
The resulting compressed expansion will contain one term for each
element of $V$, and therefore the optimal compression is that for
which $V$ has the fewest elements.

Finding the minimal set of sets $V$ that covers a set $P$ is
otherwise known as the set-covering problem.
This can be expressed as a binary linear program
\cite{Beasley_set_cover}, that is, an optimization problem where a
linear objective function $f$ of the binary unknowns $\{x_i\}$ is to
be optimized subject to a set of linear equalities and/or
inequalities involving the unknowns.
In the binary linear program associated with a set-covering problem
there are $|U|$ unknowns $\{x_i\}$ that determine whether a subset
$u_i$ is present in $V$ ($x_i=1$) or not ($x_i=0$).
The objective function that must be minimized is the number of subsets
in $V$,
\begin{equation}
\label{eq:LPObjective}
f(\{x_i\}) = \sum_{i=1}^{|U|} x_i \;,
\end{equation}
constrained so that each element in $P$ appears exactly once in $V$,
\begin{equation}
\label{eq:LPConstraints}
\sum_{j=1}^{|U|} a_{ij} x_j = 1 \;,
\end{equation}
where the binary element $a_{ij}$ of the constraint matrix indicates
whether the term $p_i$ is contained within subset $u_j$.

The size of the enumeration set $U$ can become very large if large
sets of combinable determinants are present in the original
expansion, since all possible combinations of those determinants are
required to be individual elements of $U$.
In practice we construct a set $W$ which only contains the sets in $U$
that are either of size one or not wholly contained in another set.
The maximum size of $W$ is linear with the original expansion size.
The process of constructing $V$ from $W$ differs slightly from that
described earlier, in that when an element $w_i$ is added to $V$ we
now require that the terms contained in $w_i$ be removed from all
other $\{w_j\}_{j \neq i}$.
Note that by construction the order in which elements of $W$ are
added to $V$ can affect which compression operations are in $V$, but
not their number, and therefore the compressed expansion obtained by
this procedure is of the same size as that obtained directly from
$U$.

The linear program to be solved in this simplified variation of the
method has $|W|$ unknowns $\{y_i\}$ that determine whether a subset
$w_i$ should be added to $V$.
The objective function that is to be minimized is
\begin{equation}
\label{eq:LPObjective_W}
g(\{y_i\}) = \sum_{i=1}^{|W|} y_i \;,
\end{equation}
and the constraints which guarantee that $V$ covers $P$ are
\begin{equation}
\label{eq:LPConstraints_W}
\sum_{j=1}^{|W|} b_{ij} y_j \geq 1 \;,
\end{equation}
where the binary element $b_{ij}$ of the constraint matrix indicates
whether the term $p_i$ is contained within subset $w_j$.

We further reduce the size of the linear program by partitioning $W$
into subsets such that each term $p_i$ appears in only one of the
subsets.
Solving the linear programs for each of the partitions independently
is equivalent to solving the linear program for $W$.

There exist efficient methods to solve binary linear programs, such
as the iterative simplex method implemented in the \textsc{lpsolve}
library \cite{lpsolve}.
However, solving a binary linear program, or otherwise solving the
set-covering problem, is in general NP-hard \cite{Karp_combinatorial},
and therefore a good determinant compression algorithm should
implement an approximate fall-back method for cases where it is
infeasible to obtain the exact solution in a reasonable amount of
time.
A good approximate solution to the set-covering problem can be found
in polynomial time using a ``greedy'' algorithm
\cite{greedy_setcover}, in which $V$ is constructed by adding to it
the largest element of $W$, removing all the terms contained in this
element from the other elements of $W$, and repeating this process
until no non-empty elements remain in $W$.
In the examples we have studied in the present work we have not found
any cases where we had to resort to the greedy algorithm.

\subsection{Multiple Iterations}
\label{sec:multi_iter}

In some cases it is possible for a set of compressed determinants to
be combined in order to yield an even shorter expansion.
For example the sum
\begin{equation}
\label{eq:multigroupexample_1}
[ \phi_{a_1}, \phi_{a_3} ] + [ \phi_{a_2}, \phi_{a_3} ] +
[ \phi_{a_1}, \phi_{a_4} ] + [ \phi_{a_2}, \phi_{a_4} ] \;,
\end{equation}
can be compressed into
\begin{equation}
\label{eq:multigroupexample_2}
[ \phi_{a_1}+\phi_{a_2}, \phi_{a_3} ] +
[ \phi_{a_1}+\phi_{a_2}, \phi_{a_4} ] \;,
\end{equation}
which can be further compressed into
\begin{equation}
\label{eq:multigroupexample_3}
[ \phi_{a_1}+\phi_{a_2}, \phi_{a_3}+\phi_{a_4} ] \;.
\end{equation}
The presence of multiply compressible sets of terms is a property of
the original expansion.
Note that the result of applying a compression operation to
already-compressed determinants continues to be of the form given by
Eqs.\ (\ref{eq:compressed_exp}), (\ref{eq:compressed_coeff}), and
(\ref{eq:compressed_orb}).

The most straightforward method of dealing with multiply compressible
terms is to apply the procedure described in the previous section
iteratively until no further decrease in the length of the expansion
occurs, which we refer to as the ``simple iterative method''.
By construction this method will operate on a different row of the
determinants at each iteration, and therefore the maximum number of
iterations is the number of electrons in the system.

However, the simple iterative method is not guaranteed to give the
optimal solution for two reasons.
Firstly, the choice of which terms are grouped in earlier iterations
affects the size of the final expansion, in such a way that making
sub-optimal choices at individual iterations (e.g., using the greedy
algorithm) may yield a better overall compression than solving the
set-covering problem exactly at all iterations.
And secondly, terms in the original expansion should be allowed to
contribute to more than one term of the compressed expansion.
For example, consider the following compression of a six term
expansion into two terms,
\begin{eqnarray}
\nonumber
& [ \phi_{a_1}, \phi_{a_2} ] + 2 [ \phi_{a_1}, \phi_{a_3} ] +
  [ \phi_{a_2}, \phi_{a_3} ] + & \\ \nonumber
& [ \phi_{a_1}, \phi_{a_4} ] + 2 [ \phi_{a_2}, \phi_{a_4} ] +
  [ \phi_{a_3}, \phi_{a_4} ] = & \\ \nonumber
& [ \phi_{a_1}, \phi_{a_2}+\phi_{a_3} ] +
  [ \phi_{a_1}+\phi_{a_2}, \phi_{a_3} ] + & \\ \nonumber
& [ \phi_{a_1}+\phi_{a_2}, \phi_{a_4} ] +
  [ \phi_{a_2}+\phi_{a_3}, \phi_{a_4} ] = & \\
\label{eq:example_complex_multi_1}
& [ \phi_{a_1}-\phi_{a_4}, \phi_{a_2}+\phi_{a_3} ] +
  [ \phi_{a_1}+\phi_{a_2}, \phi_{a_3}+\phi_{a_4} ] \;. &
\end{eqnarray}
This compression operation is possible only if the second and fifth
terms of the original expansion are used twice; otherwise the result
would be a three-term compressed expansion at best.
Note that, in the absence of multiply compressible terms, the resulting
compressed expansion will contain the same number of terms regardless
of whether a term can be used more than once or not.

We have developed a multiple iteration algorithm that solves the first
of these issues, although not the second which would require an
entirely different methodology, and in our opinion the resulting
method would not give significantly better compression ratios.
This method, which we refer to as the ``unified iteration method'',
is similar to that outlined in Sec.\ \ref{sec:group_terms}.
First the enumeration set $U$, which we refer to as $U^{(0)}$ in this
context, is constructed.
We define $u_i^{(1)}$ as a subset of $U^{(0)}$ such that its elements
can be combined, and $U^{(1)}$ is the set of all possible
such sets, $U^{(1)}=\{u_i^{(1)}\}$.
A similar set can be defined for each recursion level $n>1$, so that
$U^{(n)}$ is formed by all possible sets of elements of $U^{(n-1)}$
that can be combined together.
Recursion stops at $n=n_{\rm max}$ if $U^{(n_{\rm max})}$ does not
contain any terms that can be combined together.

The unknowns of the linear program for the unified iteration method
are $\{x_i^{(n)}\}$, where $x_i^{(n)}$ indicates whether set
$u_i^{(n)}$ is in $V$ or not.
The objective function that is minimized is the number of sets in $V$,
\begin{equation}
\label{eq:LPObjective_multi}
f(\{x_i^{(n)}\}) = \sum_{n=0}^{n_{\rm max}}
  \sum_{i=1}^{|U^{(n)}|} x_i^{(n)} \;,
\end{equation}
constrained by
\begin{equation}
\label{eq:LPConstraints_multi}
\sum_{n=0}^{n_{\rm max}} \sum_{j=1}^{|U^{(n)}|}
  a_{ij}^{(n)} x_j^{(n)} = 1 \;,
\end{equation}
where the binary element $a_{ij}^{(n)}$ of the constraint matrix
indicates whether or not the term $p_i$ is present in $u_j^{(n)}$.

As in the case of the simple iterative method, it is possible to
avoid constructing the enumeration set $U^{(0)}$ and instead construct
a set $W^{(0)}$ that contains all elements of $U^{(0)}$ that are not
contained in other elements.
However this simplification cannot be applied to higher recursion
levels, and one must construct $U^{(n)}$ explicitly for $n>0$.
The reason for this is that eliminating a single term $p_i$ from all
$w_j^{(n)}$ during the construction of $V$ may cause the compression
operation represented by $w_j^{(n)}$ to become invalid in the absence
of $p_i$, an event which is not taken into account by the linear
program.
Therefore the simplified linear program has the unknowns $\{y_i\}$ and
$\{x_i^{(n)}\}_{n=1}^{n_{\rm max}}$, and the objective function
\begin{equation}
\label{eq:LPObjective_multi_W}
f(\{y_i\}, \{x_i^{(n)}\}) =
 \sum_{i=1}^{|W^{(0)}|} y_i +
 \sum_{n=1}^{n_{\rm max}} \sum_{i=1}^{|U^{(n)}|} x_i^{(n)} \;,
\end{equation}
constrained so that each term of the original expansion appears at
least once in the selected operations,
\begin{equation}
\label{eq:LPConstraints_multi_W_1}
\sum_{j=1}^{|W^{(0)}|} b_{ij} y_j +
\sum_{n=1}^{n_{\rm max}} \sum_{j=1}^{|U^{(n)}|}
  a_{ij}^{(n)} x_j^{(n)} \geq 1 \;,
\end{equation}
and each term of the original expansion appears at most once
in operations of recursion level $n>0$,
\begin{equation}
\label{eq:LPConstraints_multi_W_2}
\sum_{n=1}^{n_{\rm max}} \sum_{j=1}^{|U^{(n)}|}
  a_{ij}^{(n)} x_j^{(n)} \leq 1 \;.
\end{equation}
Operations of recursion level $n>0$ must be added to $V$ before
those with order $n=0$ to prevent the application of the latter from
invalidating the former, as mentioned earlier.

Partitioning can be also applied at recursion level $n=0$ to reduce
the potential cost of solving the linear program.

\section{Implementation}
\label{sec:implement}


\begin{table*}[hbt!]
  \begin{tabular}{l @{~~~} rrr @{~~~}
      rrr@{.}l @{~~~} rrr@{.}l @{~~~} rrr@{.}l @{~~~} rrr@{.}l}
    \hline\hline
      & \multicolumn{3}{c}{Original}
      & \multicolumn{4}{c}{De-duplicate}
      & \multicolumn{4}{c}{``Quick''}
      & \multicolumn{4}{c}{``Good''}
      & \multicolumn{4}{c}{``Best''} \\
      & $N_{\rm CSF}$ & $N_s$ & $M_s$
                      & $N_d$ & $M_d$ & \multicolumn{2}{c}{$t_d$ (s)}
                      & $N_q$ & $M_q$ & \multicolumn{2}{c}{$t_q$ (s)}
                      & $N_g$ & $M_g$ & \multicolumn{2}{c}{$t_g$ (s)}
                      & $N_b$ & $M_b$ & \multicolumn{2}{c}{$t_b$ (s)} \\
    \hline
Be$_2$ &  61 &   200 &   12
             &   200 &   12 &  0&001(1) &  100 &   60 &  0&011(1)
             &    97 &   72 &  0&022(1) &   97 &   73 &  0&023(1) \\
N      &  50 &  1271 &   51
             &   764 &   51 &  0&012(1) &  324 &  191 &  0&038(2)
             &   324 &  188 &  0&044(1) &  324 &  188 &  0&054(1) \\
O      & 100 &  3386 &   53
             &  1271 &   53 &  0&058(1) &  535 &  365 &  0&125(1)
             &   534 &  361 &  0&144(1) &  534 &  361 &  0&143(1) \\
Li     & 500 &  8140 &  105
             &  5824 &  105 &  0&332(6) & 1226 & 1023 &  1&270(2)
             &  1210 & 1036 &  1&430(3) & 1210 & 1046 &  3&144(3) \\
B      & 500 & 14057 &  105
             &  5703 &  105 &  0&79(1)  &  530 &  629 &  1&802(2)
             &   529 &  629 &  1&884(7) &  529 &  631 &  3&73(1)  \\
Be     & 500 & 14212 &  105
             & 10600 &  105 &  1&18(1)  & 2218 & 1174 &  4&63(1)
             &  2177 & 1188 &  5&17(2)  & 2163 & 1198 &  7&82(3)  \\
Ne     & 400 & 22827 &  105
             & 16260 &  105 &  4&56(3)  & 1844 & 1182 & 14&79(8)
             &  1805 & 1191 & 15&04(6)  & 1805 & 1191 & 18&77(4)  \\
F      & 600 & 57456 &  105
             & 17174 &  105 & 11&43(1)  & 2801 & 2553 & 22&3(2)
             &  2749 & 2622 & 22&5(2)   & 2747 & 2627 & 24&18(4)  \\
    \hline\hline
  \end{tabular}
  \caption{
    Number of terms $N$ and number of orbitals $M$ in the original
    (sub-index $s$), de-duplicated ($d$), and compressed expansions at
    the different operational levels of the compression algorithm
    ($q$, $g$, and $b$), along with CPU time $t$ taken by the
    compression algorithm on a modest CPU.
    \label{table:results}}
\end{table*}

We have implemented the multi-determinant expansion compressor
as a stand-alone utility \cite{compress_code}
which can be used with any suitably modified quantum Monte Carlo code,
and we have modified the \textsc{casino} code \cite{casino} to be able
to use compressed multi-determinant expansions produced by the
utility.

The compression utility uses the \textsc{lpsolve} library
\cite{lpsolve} to solve the linear programs described in
Secs.\ \ref{sec:group_terms} and \ref{sec:multi_iter}.
This utility implements four operational levels:
(a) ``de-duplicate'', which performs the de-duplication stage of the
compression only (polynomial time);
(b) ``quick'', which performs de-duplication and uses the greedy
algorithm to find an approximate solution to the set-covering problems
posed by the simple iterative method (polynomial time);
(c) ``good'', which performs de-duplication and solves the
set-covering problems posed by the simple iterative method exactly
using a linear program (potentially NP-hard), and
(d) ``best'', which performs de-duplication and solves the
set-covering problem posed by the unified iteration method exactly
using a linear program (potentially NP-hard).

The ``best'' operational level should be used whenever feasible.
The ``good'' mode is provided for cases where the construction of the
enumeration sets $U^{(n)}$ makes the ``best'' mode too expensive, and
the ``quick'' mode is useful for expansions which make the ``good''
algorithm exhibit its potential NP-hardness.
The utility specifies the compressed expansion in terms of the
variables introduced in Eqs.\ (\ref{eq:compressed_exp}),
(\ref{eq:compressed_coeff}), and (\ref{eq:compressed_orb}).

We have modified the \textsc{casino} code to enable it to compute the
compressed orbitals as appropriate linear combinations of the original
orbitals, as per Eq.\ (\ref{eq:compressed_orb}).
During optimization, the original expansion coefficients are exposed
to the optimizer, and the linear coefficients of the compressed
determinants and orbitals are re-evaluated when the parameters change.

\section{Results}
\label{sec:results}

We have tested our compression algorithm on multi-determinant
expansions for the N, O, Li, B, Be, Ne, and F atoms generated using
the \textsc{atsp2k} multi-configurational Hartree-Fock (MCHF)
package \cite{atsp2k}, and with a multi-determinant expansion for the
Be$_2$ molecule generated using the \textsc{gamess} code
\cite{gamess}.
The test cases are intended to represent realistic calculations, such as
the multi-determinant calculations of Seth \textit{et
al.}\ \cite{Seth_2011_atoms}.

The wave functions produced by \textsc{atsp2k} are arranged in CSFs,
each of which comprises a set of determinants with a certain symmetry.
Since the same determinant product may be contained in different CSFs,
the atomic wave functions in our tests benefit from de-duplication.
In contrast, the \textsc{gamess} code performs de-duplication
internally, and as a result the de-duplication stage of our
compression utility does not yield any gains for the Be$_2$ wave
function.

The results for the compression algorithm are presented in Table
\ref{table:results}, which gives the number of determinants and
distinct orbitals for the original and compressed expansions at
the different operational levels of the compression utility.
Also given is the CPU time taken by the compression utility, averaged
over 20 trials, on a modest CPU.
(The compression tests were performed on a single core of a 2007
Intel Core 2 Quad CPU.)

De-duplication results in significant reductions in the size of the
multi-determinant expansions for the atoms, with $N_s/N_d$ ranging
between 1.0 and 3.3.
The compression stage provides an even greater reduction, with
values of $N_d/N_b$ ranging between 2.1 and 10.8.
In total, the compression utility yields compression factors of up to
$N_s/N_b = 26.57$.

The different compression levels ``quick'', ``good'', and ``best''
yield very similar compression sizes, with ``good'' giving a small
improvement over ``quick'' of up to $N_q/N_g = 1.024$, and ``best''
yielding a smaller change over ``good'' of $N_g/N_b = 1.005$ at most.
The CPU times required by the three compression levels are also very
similar, which indicates that the linear programs do not exhibit their
potential NP-hardness in any of the cases studied here.

The number of distinct orbitals in the compressed expansions appears
to be roughly of the order of the number of terms in the compressed
expansion, $M \sim N$.
For a given system, $M$ can be expected to increase as $N$ decreases.
However, this is not always true in our tests, e.g., for Be
$M_g < M_q$ even though $N_g < N_q$, and for Ne $M_b < M_g$ even
though $N_b = N_g$.
These cases are allowed by construction, since our algorithms do not
attempt to minimize $M$, and the different operational levels
might pick different orbital groupings that yield the same value of
$N$ but different values of $M$.

\begin{table}
  \begin{tabular}{l r r@{.}l r r@{.}l}
    \hline\hline
    & $N_s/N_b$ & \multicolumn{2}{c}{$T_s/T_b$}
    & $N_d/N_b$ & \multicolumn{2}{c}{$T_d/T_b$} \\
    \hline
    Be$_2$ &  2.06 ~ &  1&885(3) &  2.06 ~ &  1&905(6) \\
    N      &  3.92 ~ &  3&718(9) &  2.36 ~ &  2&276(8) \\
    O      &  6.34 ~ &  6&09(1)  &  2.38 ~ &  2&309(5) \\
    Li     &  6.73 ~ &  6&50(2)  &  4.81 ~ &  4&64(1)  \\
    B      & 26.57 ~ & 25&23(4)  & 10.78 ~ & 10&05(2)  \\
    Be     &  6.57 ~ &  6&48(1)  &  4.90 ~ &  4&83(2)  \\
    Ne     & 12.65 ~ & 13&17(2)  &  9.01 ~ &  9&34(1)  \\
    F      & 20.92 ~ & 21&77(5)  &  6.25 ~ &  6&48(1)  \\
    \hline\hline
  \end{tabular}
  \caption{
    Speed-up provided by the ``best'' compression algorithm over the
    uncompressed ($T_s/T_b$) and de-duplicated ($T_d/T_b$) expansions
    for a fixed number of moves in a multi-determinant VMC
    calculation.
    The relative expansion sizes $N_s/N_b$ and $N_d/N_b$ are also
    shown for comparison.
    \label{table:CASINOcputimes}}
\end{table}

To test the benefits of using compressed multi-determinant expansions
in QMC calculations, we have run multi-determinant VMC calculations
using the original, de-duplicated and ``best'' expansions.
The CPU time $T$ taken by these runs, averaged over 10 trials, is
compared in Table \ref{table:CASINOcputimes}.

In principle, the cost $T$ of the QMC calculation is at most
proportional to the expansion size $N$, and thus
$T_s/T_b \leq N_s/N_b$.
The results for both the F and Ne atoms are anomalous since this
inequality does not hold.
Our interpretation is that this is an effect of the reduced memory
footprint of the compressed expansion, for which the CPU caches can
hold the numerical data for the entire expansion and carry out the
operations more rapidly.

The data for the other systems show that the value of $T_s/T_b$ is
between 91\% and 99\% that of $N_s/N_b$, implying that handling the
determinants in the expansion is by far the leading contribution to
the CPU time of the QMC calculation, and that reducing $N$ produces an
almost equal reduction in $T$.
The additional CPU time required to compute the orbitals for the
compressed expansion as linear combinations of the original orbitals
has an insignificant impact on the benefits of using the compression
scheme for the examples in Table \ref{table:CASINOcputimes}.

We have also run VMC and DMC calculations using Jastrow factors and
backflow transformations with identical conclusions, but we have
omitted the results from Table \ref{table:CASINOcputimes} for the
sake of conciseness.

\begin{figure}[h]
  \begin{center}
    \includegraphics[width=0.40\textwidth]{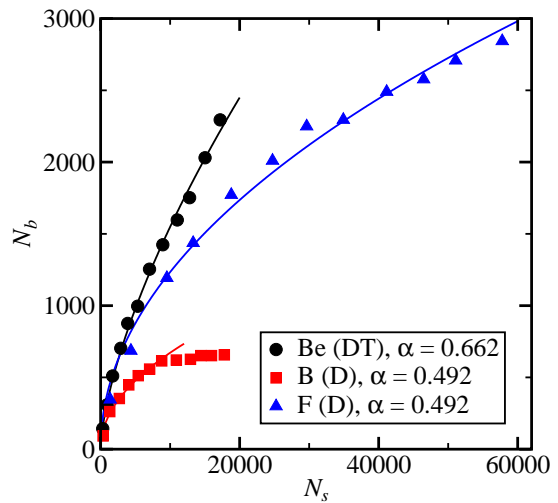}
    \caption{(Color online)
             Compressed expansion size $N_b$ as a function of the
             original expansion size $N_s$ for the Be, B, and F atoms.
             The active space used in the generation of the MCHF
             expansion includes up to double excitations (D) for B and F,
             and up to triple excitations (DT) for Be.
             The results, ignoring the plateau in the case of B, were
             fitted to $N_b = a N_s^\alpha$.
             \label{fig:detscale_overview}}
  \end{center}
\end{figure}

\begin{figure}[h]
  \begin{center}
    \includegraphics[width=0.40\textwidth]{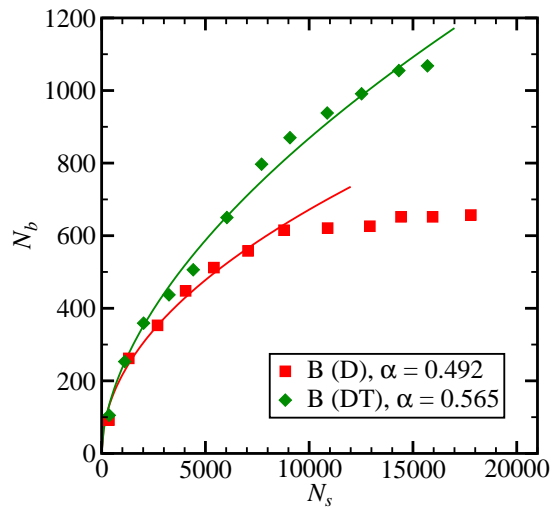}
    \caption{(Color online)
             Compressed expansion size $N_b$ as a function of the
             original expansion size $N_s$ for the B atom when the
             active space used in the generation of the MCHF
             expansion includes up to double excitations (D) and up to
             triple excitations (DT).
             The results, ignoring the plateau in the case of B (D),
             were fitted to $N_b = a N_s^\alpha$.
             \label{fig:detscale_boron}}
  \end{center}
\end{figure}

We have applied our ``best'' compression algorithm to expansions of
different sizes to investigate how $N_b$ varies with $N_s$.
Results for the Be, B, and F atoms with up to 600 CSFs are plotted in
Fig.~\ref{fig:detscale_overview}, where we also show fits to
$N_b = a N_s^\alpha$.
The active space used in the generation of the MCHF wave function
included up to double excitations for B and F, and up to triple
excitations for Be.
We find that $\alpha$ is about $1/2$ for B and F, and about $2/3$
for Be.

In the case of B we detect a plateau in $N_b$ as a function of $N_s$.
We interpret this as a sign of the exhaustion of the finite active
space used in the generation of the multi-determinant wave function.
To test this hypothesis we have repeated the scaling tests for
B with an active space that includes up to triple excitations,
as shown in Fig.~\ref{fig:detscale_boron}.
As we expected, the plateau disappears when the larger active space
is used.

We conclude that the use of the compression algorithm can make
multi-determinant QMC calculations scale as $N_s^\alpha$, with
$1/2 \leq \alpha < 1$.

\section{Conclusion}
\label{sec:conclude}

In this paper we have presented a compression algorithm for
multi-determinant expansions based on a simple identity for combining
determinants, which can provide computational cost savings in QMC
calculations of about the compression factor $N_s/N_b$, which in our
tests ranges between 2.06 and 26.57.

In addition to the full compression algorithm we have implemented a
polynomial-scaling fall-back algorithm which has been shown to yield
nearly the same compression ratios as the full method.
This algorithm avoids the NP-hardness associated with solving the
set-covering problem, but in none of our tests did the full algorithm
incur costs excessive enough to require the use of the fall-back
algorithm.

We find that the compression algorithm makes QMC calculations scale
sub-linearly with the number of determinants in the expansion.
The cost savings provided by using compressed determinant expansions
are expected to permit QMC calculations using much larger
multi-determinant expansions than would otherwise be possible.
Our compression algorithm can be used in combination with methods for
the optimized evaluation of multi-determinant wave functions
\cite{clark_2011,nukala_2009} for additional efficiency.

\begin{acknowledgments}
We thank John Trail for producing the multi-determinant wave function
for Be$_2$.
The authors acknowledge financial support from the Engineering and
Physical Sciences Research Council (EPSRC) of the United Kingdom.
Computational resources were provided by the Cambridge High Performance
Computing Service.
\end{acknowledgments}








\end{document}